\documentclass[namedreferences]{kluwer}
\usepackage{graphicx}

\providecommand{\atanh}{\mathrm{atanh}}
\renewcommand{\a}{\alpha}
\renewcommand{\b}{\beta}

\begin{document}

\begin{article}
\begin{opening}    
  
  \title{Modeling Society with Statistical Mechanics:\\ an
    Application to Cultural Contact and Immigration}
    
  \author{Pierluigi \surname{Contucci}\email{contucci@dm.unibo.it}}

  \institute{Department of Mathematics, University of Bologna}

  \author{Stefano \surname{Ghirlanda}\email{stefano.ghirlanda@unibo.it}}
    
  \institute{Department of Psychology, University of Bologna\\
    Group for Interdisciplinary Cultural Studies, Stockholm
    University}

  \runningauthor{P. Contucci, S. Ghirlanda}

  \runningtitle{Statistical mechanics of cultural contact}

  \begin{abstract}
    We introduce a general modeling framework to predict the outcomes,
    at the population level, of individual psychology and behavior.
    The framework prescribes that researchers build a cost function
    that embodies knowledge of what trait values (opinions, behaviors,
    etc.) are favored by individual interactions under given social
    conditions. Predictions at the population level are then drawn
    using methods from statistical mechanics, a branch of theoretical
    physics born to link the microscopic and macroscopic behavior of
    physical systems.  We demonstrate our approach building a model of
    cultural contact between two cultures (e.g., immigration), showing
    that it is possible to make predictions about how contact changes
    the two cultures.
  \end{abstract}
  
\end{opening}           

\section{Introduction}
\label{sec:intro}

Modern society features an increasing degree of interaction between
cultures (``cultural contact'') owing to, e.g., communication
technologies, immigration and other socio-political forces
\cite{lull00}. In many countries cultural contact is perceived as both
an opportunity and a threat and is related to important public issues
such as immigration management and the need to ``protect'' national
culture \cite{corn94,give05}. Our understanding of these phenomena is,
however, limited: we cannot predict the outcome of cultural contact,
nor make plausible conjectures that can be used in policy making.
Within this context, the aim of this paper is twofold: we first
describe a general mathematical framework for modeling social
interactions, then we make specific assumptions relevant to studying
immigration, i.e., social contact between two groups that, typically,
differ both in culture and relative size.

\section{General framework}
\label{sec:general}

\subsection{A formalization of social interactions}
\label{sec:formalization}

For simplicity, we focus on a single \textit{cultural trait}, which
may represent an idea, an opinion or a behavior that has two mutually
exclusive forms.  A useful example to keep in mind is being in favor
or against an issue such as the death penalty, or any other issue that
might be the subject of a yes/no referendum vote. Our framework allows
to consider multiple traits without conceptual differences, although
model analysis may in general be much more difficult.

We consider a population of $N$ individuals, labeled by an index $i$
ranging from $1$ to $N$.  We associate to the $i$-th individual the
variable $s_i$, which may take values $+1$ or $-1$ representing the
two possible trait values.  For instance, $+1$ might represent a yes
vote in a referendum, and $-1$ a no vote. The state of the whole
population is thus encoded in an array $s$ of $N$ numbers, such as
$s=\{+1, -1, +1,\ldots\}$.

The hallmark of social interactions is that individuals may change
their opinions or behavior owing to interactions with others. A given
couple $(i,k)$ can in principle be in one of the four states
$\{+1,+1\}$, $\{-1,+1\}$, $\{+1,-1\}$ and $\{-1,-1\}$, but these
outcomes, in general, do not have the same probability. Which one is
more likely will depend on the characteristics of individuals such as
their culture and personality.  Our starting assumption is that
individuals have no particular bias towards $+1$ or $-1$ opinions:
what matters most is whether, by adopting one or the other value, an
individual is in agreement or disagreement with others.  There are two
reasons for this assumptions. First, social psychologists have shown
that, in most cultures, agreement or disagreement with others is a
powerful determinant of individual opinions and behavior, often more
important than holding a particular opinion \cite{bond96}; we will
expand on this point in our model of immigration below.  Second, our
framework allows to introduce biases that favor a particular trait
value, if needed. Indeed, any model in which individuals are biased
can be recast as a model with unbiased individuals, plus an additional
``force'' that orients individual opinions. Thus our starting
assumption of unbiased individuals does not reduce the generality of
the framework.  Again, we will make a specific example for the case of
immigration below.

To formalize these notions, we assume that individuals take on the
trait that minimizes a \textit{cost function}.  We define the cost
$H_{ik}$ for individual $i$ to agree or disagree with individual $k$
as
\begin{equation}
  \label{eq:Hik}
  H_{ik}(s_i,s_k)=-J_{ik}s_is_k  
\end{equation}
where $J_{ik}$ is a number that summarizes the nature of the
interaction between $i$ and $k$, as follows. When $i$ and $k$ agree
($s_is_k=1$) we have a cost $H_{ik}=-J_{ik}$, while when $i$ and $k$
disagree ($s_is_k=-1$) we have $H_{ik}=J_{ik}$. Thus whether
agreement or disagreement carries the lesser cost depends on the sign
of $J_{ik}$: $J_{ik}>0$ favors agreement while $J_{ik}<0$ favors
disagreement. The magnitude of $J_{ik}$ gives how important it is for
$i$ to agree or disagree with $j$.  If, for instance, $J_{ik}>J_{im}$
then it is more important for $i$ to agree with $k$ than with $m$,
while $J_{ik}=0$ means that agreement with $k$ is not relevant to $i$.
The signs and magnitudes of the $J_{ik}$'s become important when we
consider that an individual interacts with many others. In this case,
we assume that the costs $H_{ik}$ relative to each interaction sum up
to a total cost $H_i$
\begin{equation}
  \label{eq:Hi}
  H_i(s_i)= \sum_k H(s_i,s_k)=-\sum_k J_{ik}s_is_k
\end{equation}
As anticipated above, we can take into account additional factors that
may influence individuals modifying equation (\ref{eq:Hi}) as follows:
\begin{equation}
  \label{eq:Hh}
  H_i(s_i) = -\sum_{k}J_{ik}s_is_k - h_is_i  
\end{equation}
meaning that individual $i$ is subject to an additional ``force'' that
favors $s_i=1$ if $h_i>0$ and $s_i=-1$ if $h_i<0$. The quantity
$h_i$ may represent any factor that is not explicitly taken into
account by the direct interaction with other individuals. For
instance, it may summarize the influence of media, government
campaigns or existing culture (see below).

We can now write a population-level cost function as the sum of
individual cost functions:
\begin{equation}
  \label{eq:H}
  H(s) = \sum_i H_i(s_i) = -\sum_{i,k}J_{ik}s_is_k -\sum_ih_is_i
\end{equation}
We stress that the cost function is a theoretical computational aid to
track which trait values are favored by the interactions $J_{ik}$ and
the external forces $h_i$. We do not assume that individuals
explicitly compute or are aware of such costs. Rather, $H(s)$ should
be designed so that its minima correspond to those trait values that
are favored by the mechanisms with which individuals interact.

\subsection{The role of statistical mechanics}
\label{sec:statmech}

Once a cost function has been specified, it is possible to calculate
population level quantities such as the average trait value using the
methods of statistical mechanics, a branch of theoretical physics.
Statistical mechanics was originally concerned with deriving the laws
of thermodynamics from the behaviour of atoms and molecules \cite{th},
but can actually be applied to understand the macroscopic (population
level) properties of any system composed of many parts that interact
according to given microscopic (individual level) rules. More recently
its methods have found application in fields as diverse as biology
\cite{meza87}, neuroscience \cite{amit89,arbi03}, economy and finance
\cite{bp} and also social science \cite{durl99}. The starting point is
to assign to each system configuration $s$ a probability $\Pr(s)$
according to the Boltzmann-Gibbs formula \cite{th}
\begin{equation}
  \label{eq:bg}
  \Pr(s) = \frac{e^{- H(s)}}{\sum_s e^{- H(s)}}
\end{equation}
where the sum runs over all possible configurations of the
system.\endnote{In physical applications, the parameter $\b$ (inverse
  temperature) usually multiplies $H$ in (\ref{eq:bg}). Here, as in
  other applications of statistical mechanics such as combinatorial
  optimization \cite{meza87}, we include $\b$ in $H$ itself as an
  overall scale factor.}

By means of (\ref{eq:bg}) a given configuration is considered more or
less likely according to whether it is more or less costly: a low
value of $H(s)$ results in a high probability of $s$, and vice-versa.
Assigning probabilities based on a given cost function is the heart of
statistical mechanics and is inspired by the principles of
thermodynamics (see the appendix for a short discussion, and
\opencite{th}, for a fuller treatment).

Once a probability is assigned to system configurations, it is
possible to compute the expected values of quantities of interest and
to relate them to the parameters that describe the system. For
instance the average cultural trait defined by
\begin{equation}
  m(s)=\frac{1}{N}\sum_{i}s_i
\end{equation}
would have an expected value given by
\begin{equation}
  \label{eq:av}
  \overline{m} = \sum_s m(s) \Pr(s)
\end{equation}
Note that, while $m(s)$ is the average trait value in a given
configuration, $\overline m$ is the average trait value over all
possible system configurations, each one weighted according to its
probability of occurrence.  These probabilities depend on the cost
function $H$ and thus on the parameters that appear in its expression,
i.e., the $J_{ik}$'s and $h_i$'s.

Rather than directly attempting to compute expected values such as
(\ref{eq:av}), statistical mechanics aims to compute the so-called
\textit{free energy} of a system, defined as
\begin{equation}
  \label{eq:fe}
  f = - \log \sum_s e^{- H(s)}
\end{equation}
The rationale for this strategy is that important quantities such as
(\ref{eq:av}) can be easily computed from knowledge of the free energy
function, typically by taking suitable derivatives with respect to
system parameters (see appendix).  The basic task of statistical
mechanics is thus, after the cost function $H$ has been specified, to
calculate the free energy function for a given system. We stress that
\textit{the form of the cost function is not given by statistical
  mechanics; rather, it is the outcome of a modeling effort relative
  to a specific problem.}  We now make an example of how one may
proceed.

\section{The cultural outcomes of immigration}
\label{sec:immigration}

We illustrate here the potentials of our framework considering the
impact of immigration on culture. We consider a large and a small
population, which will be referred to, respectively, as
\textit{residents} ($R$) and \textit{immigrants} ($I$).  We let $N_1$
be the number of residents, and $N_2$ of immigrants, with $N_1\gg N_2$
and $N=N_1+N_2$ the total number of individuals. We are interested in
how cultural contact changes the average trait values in the two
populations, with the aim of understanding the effect of one culture
upon the other.

\subsection{Interactions between residents and immigrants}
\label{sec:interactions}

Our main assumption regarding how residents and immigrants interact is
that people, generally speaking, tend to agree with those who are
perceived as similar to oneself and to disagree with those perceived
as different. In social psychology this is known as the
\textit{similarity-attraction} hypothesis. It has received ample
support, although the details of how we interact with others often
depend on social context \cite{mich02,byrn97}. We consider this
assumption a general guideline, and in modeling a specific case it can
be modified without difficulty. We formalize the similarity-attraction
hypothesis by assuming that high perceived similarity corresponds to
positive values of $J_{ik}$, and low perceived similarity to negative
values. Since residents and immigrants have generally different
cultures, we may assume the following structure for the interaction
coefficients $J_{ik}$. We let the interaction between any two
residents be $J^{(1,1)}>0$; the similarity-attraction hypothesis
suggests that this be a positive number, whose magnitude reflects how
strongly residents prefer to agree among themselves. Likewise, we let
the interactions between immigrants be $J^{(2,2)}>0$. The mutual
interactions $J^{(1,2)}$ and $J^{(2,1)}$ should model whether
residents prefer to agree or disagree with immigrants, and vice-versa,
and how strongly so. If resident and immigrant cultures are very
different, the similarity-attraction hypothesis suggests to take both
$J^{(1,2)}$ and $J^{(2,1)}$ as negative, but the best choice of values
depends on the specific case one intends to model.

Note that we are assuming that $J_{ik}$ depends only on population
membership and not on the particular individuals $i$ and $k$ (the
so-called {\it mean field} assumption in statistical mechanics).  This
assumption greatly simplifies mathematical analysis but is not wholly
realistic. It can capture the average variation in interactions across
population but not the variation that exists within each population.
For instance, a more realistic assumption would be to take the
$J_{ik}$'s as random variables whose mean and variance depend on
population membership.  We plan to return on that model (which would 
represent the two-population generalization of the Sherrington-Kirkpatrick 
model in statistical mechanics, \opencite{meza87}) in future studies.

When modeling interactions, a technical requirement is that the value
of the cost function be proportional to total population size $N$.
This guarantees that the free-energy function and important quantities
such as average trait value, equation (\ref{eq:av}), scale
appropriately with $N$. In our case the appropriate scaling is $1/2N$,
hence the interactions are:
\begin{equation}
  \label{eq:J}
  \renewcommand{\arraystretch}{2.75}
  J_{ik} = \left\{
  \begin{array}{l@{\qquad}l}
    \displaystyle\frac{J^{(1,1)}}{2N} & i,k\in R\\
    \displaystyle\frac{J^{(1,2)}}{2N} & i\in R, k\in I\\
    \displaystyle\frac{J^{(2,1)}}{2N} & i\in I, k\in R\\
    \displaystyle\frac{J^{(2,2)}}{2N} & i,k\in I
  \end{array}\right.
\end{equation}

\subsection{Modeling pre-existing culture}
\label{sec:pre-existing}

Before the two populations start to interact, residents and immigrants
are each characterized by a given average trait value, say ${\tilde
  m_1}$ and ${\tilde m_2}$, respectively. We consider ${\tilde m_1}$
and ${\tilde m_2}$ as experimental data about the beliefs or behavior
of each population, which could be obtained from, say, a referendum
vote on a particular issue (e.g., the death penalty) or from
statistical sampling of the population.

That a population is characterized by a given average value $\tilde
m_i\ne0$ means that the two forms of the trait are not equally common.
Specifically, the individuals with the $+1$ form are $N(1+\tilde
m_i)/2$, while $N(1-\tilde m_1)/2$ individuals have the $-1$ form.
Pre-existing culture, in other words, is like a bias or force that
favors one trait value over the other. For modeling purposes, it is
convenient to describe pre-existing culture as a ``cultural force''
that acts to orient the opinion of otherwise unbiased individuals.
This is possible including a force term in the cost function, as shown
in (\ref{eq:Hh}).  By standard methods of statistical mechanics (see
appendix) it is possible to show that the force term corresponding to
a particular average opinion $\tilde m_i$ is
\begin{equation}
  \label{eq:h}
  h^{(i)} = \atanh(\tilde m_i)-J^{(i,i)}\tilde m_i
\end{equation}
where $\atanh$ is the inverse hyperbolic tangent function.  To
summarize, a model in which individuals are biased so that the average
opinion is $\tilde m_i$ is equivalent to a model with unbiased
individuals subject to a force given by (\ref{eq:h}).

\subsection{Model analysis}
\label{sec:analysis}

So far we have specified interaction terms $J_{ik}$ to model cultural
contact between two populations and we have introduced equation
(\ref{eq:h}) to represent the pre-existing culture in the two
populations. The next step is to compute the average trait values
$m_1$ and $m_2$ in the two populations after immigration has taken
place. The same method that allows to derive equation (\ref{eq:h})
enables us to derive the following equations for $m_1$ and $m_2$ (see
appendix):
\begin{subequation}
  \label{eq:m1m2}
  \begin{equation}
    \label{eq:m1}
    m_1 = \tanh\Big( (1-\a)J^{(1,1)}m_1 + \a(J^{(1,2)}+J^{(2,1)})m_2 +
    h^{(1)} \Big)
  \end{equation}
  \begin{equation}
    \label{eq:m2}
    m_2 = \tanh\Big( (1-\a)(J^{(2,1)} + J^{(1,2)})m_1 + \a
    J^{(2,2)}m_2 + h^{(2)}\Big)
  \end{equation}
\end{subequation}
where $\a=N_2/N$ is the fraction of immigrants in the total population
and $\tanh$ is the hyperbolic tangent function.  The values of $m_1$
and $m_2$ predicted by (\ref{eq:m1m2}) depend of course on values of
the $J$ and $h$ parameters, and on $\a$. We give here a qualitative
description of the different regimes that one can observe varying
these parameters.  We refer to \inlinecite{co} for a proof of the
following statements, in the context of an analogous model from
condensed matter physics.

The two key parameters are $\a$, the fraction of immigrants, and the
overall scale of the interactions $J_{ik}$, which we label $J$.  If
$J$ is below a critical value $J^\star$, equation (\ref{eq:m1m2}) has
always one pair of solutions, for all values of $\a$. In this case the
two populations are essentially merging into a homogeneous one, with
average cultural trait in between the two initial ones---more toward
one or the other according to the value of $\alpha$. This regime is
not surprising and corresponds to the na\"ive prediction that one
could have made a priori without applying statistical mechanics.

If the interaction scale is large ($J>J^\star$), however, model
predictions are highly non-trivial, suggesting that the outcome of
cultural contact can be surprising.  Depending on $J$ there are two
critical values for $\alpha$: $\alpha_1(J)$ and $\alpha_2(J)$ that
delimit qualitatively different behavior.  For $\alpha\le \alpha_1(J)$
the resident culture dominates dominant and the immigrant culture
disappears, i.e., $m_2$ is close to $m_1$ irrespective of the initial
value $\tilde m_2$. The converse happens when $\alpha\ge \alpha_2(J)$,
i.e., immigrant culture dominates.  The most interesting case occurs
when $\alpha_1(J) \le \alpha \le \alpha_2(J)$. In this regime
(\ref{eq:m1m2}) has two distinct solutions in which either of the two
cultures dominates. That is, both cultures may survive the immigration
process, generally with a different probability determined by system
parameters.
\begin{figure}[t]
  \centering
  \includegraphics[angle=-90,width=.6\textwidth]{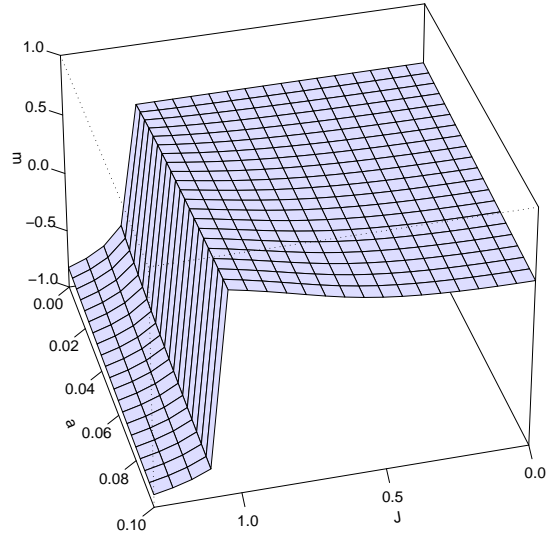}
  \caption{Possible  outcomes of cultural contact. Average trait value
    in the two populations, $m=(1-\a)m_1+\a m_2$, is plotted as a
    function of $\a$ (fraction of immigrants) and $J$ (scale of the
    interaction) for an interaction matrix of the form (\ref{eq:J})
    with $J^{(i,k)}=J$. Before the interaction the two populations
    have initial trait values of $\tilde m_1=0.5$ and $\tilde
    m_2=-0.5$. Only the most likely outcome is plotted (see text).}
  \label{fig:Jalpha}
\end{figure}

The parameter values that favor the resident or immigrant culture,
have still to be worked out and will be the topic of future
work.\endnote{As noted above, \inlinecite{co} has studied a formally
  identical model arising from a condensed-matter problem, but there
  the $h_i$'s were free parameters, while here they are determined in
  terms of the $J^{(i,i)}$ and the $\tilde{m}_i$ through
  (\ref{eq:h}).}  Here we analyze the case in which the intensity of
the interactions is uniform both within and between groups,
$J^{(i,k)}=J$.  This is interpreted as two groups that do not really
discriminate between themselves, so that disagreement with any
particular individual carries the same cost independent of which group
the individual belongs to. We assume, however, that the two groups
have initially a very different average trait value: $\tilde m_1=0.5$
and $\tilde m_2=-0.5$. In figure~\ref{fig:Jalpha} we explore this
system by plotting the average trait value after the interaction,
$m=(1-\a)m_1+\a m_2$, for $\a$ between 0 and 10\% and for
$J\ge 0$.\endnote{The maximum admissible value for $J$ is given by the
  condition that (\ref{eq:h}) has a stable solution, corresponding to
  the assumption that each culture is in equilibrium before cultural
  contact. This condition yields $J\lessapprox1.3$.}  For $J=0$ (no
interaction) $m$ is simply the weighted sum of pre-existing trait
values, $m=(1-\a)\tilde m_1+\a\tilde m_2$, where each group
contributes according to its size.  As a function of $\a$, this is a
straight line.  As the interaction increases the line slowly bends and
for higher values of $\alpha$ we see a slight exaggeration of the
pre-existing opinion $\tilde m_1$ (the surface in
figure~\ref{fig:Jalpha} rises slightly over the level $m=0.5$). When
$J$ crosses a critical value $J^\star\simeq1.125$, however, a dramatic
phenomenon occurs: the population undergoes a sudden change of
opinion, switching to a value of $m$ that is closer to, and indeed
exaggerates, the initial value in the immigrant population, $\tilde
m_2=-0.5$.  Note that this sudden transition occurs for all values of
$\alpha$, i.e., irrespective of the proportion of immigrants.  The
solution with $m$ closer to $m_1$ is still available to the system
(not plotted in Figure\thinspace\ref{fig:Jalpha}), but as $J$ grows
past $J^\star$ it is less and less likely that the system remains in
such state (technically, for $J<J^\star$ the solution with
$m\simeq\tilde m_1$ has a higher free-energy than the solution with
$m\simeq\tilde m_2$ and thus becomes metastable, allowing fluctuations
to cause a transition between the two solutions).
Thus, according to this model, to prevent dramatic changes in resident
culture, it would do little to restrict immigration (the effect of
$\a$ is small in the graph).  Rather, one should concentrate in
reducing the scale of the interaction $J$, i.e., the strength of
attitudes within and between groups.

\section{Discussion}
\label{sec:disc}

Attempts to apply mathematical-physics methods to social sciences have
appeared in the litterature since the pioneering work of
\inlinecite{ga}.  In this paper we have focused on statistical
mechanics as a tool to bridge the gap from in\-dividual-level
psychology and behavior to population-level outcomes.  Our framework
prescribes that researchers build a cost function that embodies
knowledge of what trait values (opinions, behaviors, etc.)  are
favored by individual interactions under given social conditions.  The
cost function, equation (\ref{eq:H}), is defined by a choice for the
interactions $J_{ik}$ and the fields $h_i$ that represent social
forces influencing individual opinions and behavior.  This modeling
effort, of course, requires specific knowledge of the social issue to
be modeled. After a cost function has been specified, the machinery of
statistical mechanics can be used to compute population-level
quantities and study how they depend on system parameters.

We have demonstrated our framework with an attempt to understand the
possible outcomes of contact between two cultures. Even the simple
case we studied in some detail the model suggests that cultural
contact may have dramatic outcomes (figure \ref{fig:Jalpha}).  How to
tailor our framework to specific cases, and what scenarios such models
predict, is open to future research.

\begin{acknowledgements}
  We thank F. Guerra for very important suggestions. I. Gallo, C.
  Giardina, S. Graffi and G. Menconi are acknowledged for useful
  discussion.
\end{acknowledgements}

\theendnotes

\appendix

\section{Model solution}

It is a standard result of statistical mechanics \cite{th} that the
free energy function of a system defined by a cost function of the
form
\begin{equation}
  \label{eq:HCW}
  H(s) = -\frac{J}{2N}\sum_{i,k}s_is_k -h \sum_i s_i
\end{equation}
is obtained for the value of $m$ that minimizes the function
\begin{equation}
\label{eq:fcw}
F(m)= -\frac{J}{2}m^2 -hm +\frac{1+m}{2}\log\frac{1+m}{2}
+\frac{1-m}{2}\log\frac{1-m}{2}
\end{equation} 
The minimization of this function with respect to $m$ yields the
condition (\ref{eq:h}) which relates $m$ and $h$ and the Hamiltonian
parameters.  The structure of the free energy (\ref{eq:fcw}) admits
the standard statistical mechanics interpretation as a sum of two
contributions: the internal energy (the average of the cost function)
\begin{equation}
  \label{eq:U}
  U(m)=\overline H=-\frac{J}{2}m^2-hm
\end{equation}
minus the entropy
\begin{equation}
  \label{eq:entropy}
  S(m)=-\sum_s \Pr(s)\log\Pr(s)= -\frac{1+m}{2}\log\frac{1+m}{2}
  -\frac{1-m}{2}\log\frac{1-m}{2} \; .
\end{equation}
One can indeed show that the distribution function (\ref{eq:bg}) may
be deduced from the second principle of thermodynamics i.e. as the
distribution for which the entropy is minimum at an assigned value of
the cost function \cite{th}.  Equation (\ref{eq:m1m2}) is obtained
similarly from the representation of the free energy of the two
population system as the minimum of the function
\begin{equation}
  \label{eq:fcw2}
  \renewcommand{\arraystretch}{1.5}
\begin{array}{l@{\,}l}
  F(m_1,m_2)=&-(1-\a)^2J^{(1,1)}\frac{m^2_1}{2}-\a^2J^{(2,2)}\frac{m^2_2}{2}\\
  &-\a(1-\a)(J^{(1,2)}+J^{(2,1)})m_1m_2\\
  &-(1-\a)h_1m_1-\a h_2m_2\\
&+(1-\a)[+\frac{1+m_1}{2}\log\frac{1+m_1}{2}+\frac{1-m_1}{2}\log\frac{1-m_1}{2}]\\
&+\a[+\frac{1+m_2}{2}\log\frac{1+m_2}{2}+\frac{1-m_2}{2}\log\frac{1-m_2}{2}]
  \end{array}
\end{equation} 
The minimum condition yields (\ref{eq:m1m2}).
\end{article}
\end{document}